\input epsf

\documentstyle[aps,eqsecnum]{revtex}
\title{Dissipative dynamics of fission in the framework of asymptotic
expansion of Fokker-Planck equation}
\author{Asish K. Dhara, Sailajananda Bhattacharya and Kewal Krishan}
\address{Variable Energy Cyclotron Centre, 1/AF Bidhan Nagar, Calcutta
-700064,India}
\tighten
\begin{document}

\pagenumbering{arabic}

\maketitle
\date{\today}
\begin{abstract}
The   dynamics  of  fission  has  been  formulated  by  generalising  the
asymptotic expansion of  the  Fokker-Planck  equation  in  terms  of  the
strength  of  the fluctuations where the diffusion coefficients depend on
the stochastic variables explicitly. The prescission neutron multiplicities
and mean kinetic energies of the evaporated neutrons have been calculated
and compared with the respective experimental data over a wide range of
excitation energy and compound nuclear mass. 
The mean and  the  variance  of  the
total  kinetic energies of the fission fragments have been calculated and
compared with the experimental values.

\end{abstract}

\section{introduction}
At  present,  it  is  commonly  agreed upon that the fission process is a
dissipative phenomena, where initial energy of the  collective  variables
get  dissipated  into  the  internal  degrees of freedom of nuclear fluid
giving rise to the increase in internal excitation energy. As dissipation
is referred to the interaction of the system coordinate  with  the  large
number  of  degrees of freedom of the surrounding reservoir, this process
is  always  associated  with  the  fluctuations  of   relevant   physical
observables. Thus, the dynamics of fission process resembles the standard
Brownian  motion  problem,  where  the collective variables such as shape
degrees of freedom act as 'Brownian particles' interacting stochastically
with large number of internal nucleonic degrees of  freedom  constituting
the  surrounding  'bath'.  This mesoscopic description is inevitable once
the  fluctuations  of  the  observables  are  amenable  to   experimental
observation.

There  have  been  several  attempts in the past to study the dynamics of
fission by solving either the Langevin equation \cite{abe1,abe2,frob,pomo},
or multidimensional  Fokker-Planck  equation  \cite{fp1,fp2,fp3,fp4},
which  is   a
differential  version of Langevin equation. In the case of fission, it is
experimentally observed that the variances of  the  physical  observables
are,  in  general,  small  compared  to  their  respective  mean values (
typically, the ratio of the root mean square deviation and  the  mean  of
the  kinetic energy is $\sim$ 0.1). The question naturally arises whether
one can utilise this simple fact in the  theoretical  scheme  instead  of
solving   the  Langevin  equation  (LE)  or  corresponding  Fokker-Planck
equation in detail. In this spirit, we present an alternative theoretical
prescription for the calculation  of  various  moments  of  the  physical
observables  related  to the fission process based on the assumption that
the  full  solution  of  the  Fokker-Planck  equation  (FPE)  admits   an
asymptotic  expansion  in  terms  of  strength  of  the fluctuations. The
asymptotic expansion method was first developed by van Kampen  \cite{van}
for  the  stochastic  processes  having  constant diffusion coefficients.
However, a generalisation of the above prescription is necessary  in  the
case of fission where the dissipation is usually assumed to depend on the
instantaneous  shape of the fissioning system and therefore the diffusion
coefficients are also shape dependent. To the best of our knowledge, such
an application in the case of fission is not available in the literature.

In  the  present  paper,  we  report  a  generalised  formulation  of the
asymptotic expansion of the Fokker-Planck equation  where  the  diffusion
coefficient  depends  on  the  stochastic  variables  explicitly. In this
formulation the dynamics of the stochastic processes reduces to a set  of
linear  ordinary differential equations which are far simpler to solve as
compared to either multidimensional partial differential ( Fokker-Planck)
equations or stochastic differential (Langevin) equations. Presently,  we
apply  this  formulation to calculate the various moments of the relevent
physical observables of the fission process.

The  paper  is  organised  as follows.The generalised formulation and its
application to  the  dynamics  of  fission  is  given  in  Sec.  II.  The
calculations  and  numerical  results  are discussed in Sec.III. Finally,
concluding remarks are given in Sec.IV.

\section{Asymptotic expansion of the Fokker-Planck equation}
\label{f-p}
\subsection{The Formalism}

The  mesoscopic  description  of the fission process begins with a set of
Langevin equations:

\begin{mathletters}
\begin{eqnarray}
\label{eq.1}
\dot{X_i} &=& h_i(\{X\},\{Y\}) +\eta_i(t)   \\
\dot{Y_i} &=& H_i(\{X\},\{Y\})  \hspace{18pt} ; i=1,...,N
\end{eqnarray}
\end{mathletters}

where  $h_i$  and  $H_i$ are given functions of the stochastic collective
variables $ X_1,X_2,...,X_N $ and $  Y_1,Y_2,...,Y_N  $  in  the  fission
process and $ \eta_i(t)$ refers to the driving noise term associated with
the  interaction  of the {\it i}th collective variable with the reservoir
constituting nucleonic degrees of freedom. For simplicity, we assume  the
noise  to  be a gaussian white with zero mean and decoupled for different
degrees of freedom with auto-correlation functions given by

\begin{equation}
\label{eq.2}
<\eta_i(t)> = 0,\hspace{8pt} <\eta_i(t)\eta_j(t')> = D_i(y_i)\delta(t-t')
\delta_{ij}  ,
\end{equation}

where  $D_i(y_i)$  is the diffusion coefficient associated with {\it i}th
variable,  depending  only  on  the sample space $y_i$ for the stochastic
variable $Y_i$.

The Fokker-Planck equation corresponding to the Langevin equation (2.1) is

\begin{equation}
\label{eq.3}
\frac{\partial f(\{x\},\{y\},t)}{\partial t}  =  -\sum_i [\frac{\partial
(h_i f)} {\partial x_i} + \frac{\partial (H_i  f)}{\partial  y_i}  -(1/2)
D_i(y_i) \frac{\partial^2 f}{\partial x_i^2}]   .
\end{equation}
The  quantity  $f(\{x\},\{y\},t)$  is  the  probability  density function
depending on the variables $ x_1,x_2,...,x_N, y_1,y_2,...,y_N $ and  time
$t$ explicitly. If we are interested in finding the time evolution of the
conditional  probability  distribution  function  then  we  have to solve
Eq.(2.3)   with   initial    values    $x_i(0)=x_i^0$,    $y_i(0)=y_i^0$,
$\forall$$i$,  at  $t=0$.  That  is,  we have to solve Eq.(2.3) for those
realisations which are known to start from these specific points  in  the
whole sample space.

In  the  cases  where  diffusion  coefficient  is constant the asymptotic
expansion method  of  van  Kampen  \cite{van}  consists  of  writing  the
stochastic  variables as the sum of deterministic value and a fluctuating
part at each time $t$ with root of the diffusion constant as  a  strength
of  the fluctuating part. In the present paper, we generalise this method
for the  situations  where  the  diffusion  coefficients  depend  on  the
stochastic  variables  explicitly. Such a situation is encountered in the
case  of  fission  process,  where  the  friction   coefficient   depends
explicitly  on  the  collective  variable or shape of the nucleus at each
instant of time. In this case, we further assume that, in the  asymptotic
expansion,  the  strengths  of  the  fluctuating  parts of the stochastic
variables depend only on the deterministic values of the  respective  $y$
variables:

\begin{mathletters}
\begin{eqnarray}
\label{eq.4}
x_i&=&\bar{x_i} + \surd D(\bar{y_i}) \zeta_i\\
y_i&=&\bar{y_i} + \surd D(\bar{y_i}) \xi_i
\end{eqnarray}
\end{mathletters}

The  quantities  $\{\zeta_i\},\{\xi_i\}$ refer to the fluctuations of the
stochastic variables $\{x_i\}$ and $\{y_i\}$ around  their  deterministic
values   $\{\bar{x_i}\},  \{\bar{y_i}\}$.  Next,  we  introduce  the  new
distribution   function   $Q$   depending   only   on    the    variables
$\{\zeta_i\}$,$\{\xi_i\}$  and  $t$. The normalisation condition suggests
that the $f$ and $Q$ will be related by

\begin{equation}
\label{eq.5}
f(\{x\},\{y\},t)  =  {\prod_{i=1}^N[D_i (\bar{y_i})]^{-1}} Q_i(\zeta_i,
\xi_i,t)
\end{equation}

Substituting  Eq.(2.4)  in the Fokker-Planck equation(2.3), making Taylor
expansion of $h(\{x\},  \{y\}),  H(\{x\},  \{y\})$  around  $\{\bar{x}\},
\{\bar{y}\}$   and   collecting   coefficients  of  various  order  of  $
D(\bar{y_i})$,we could generate a hierarchy of  equations.  As  expected,
the first set would give rise to the equation of motion for $\{\bar{x}\}$
and $\{\bar{y}\}$.

\begin{mathletters}
\begin{eqnarray}
\label{eq.6}
\dot{\bar{x_i}} &=& h_i(\{\bar{x}\},\{\bar{y}\})  \\
\dot{\bar{y_i}} &=& H_i(\{\bar{x}\},\{\bar{y}\}) \hspace{18pt};\forall{i}
\end{eqnarray}
\end{mathletters}

Eqs.(2.6) are the Euler-Lagrange equation for deterministic motion. These
equations  are  to  be  solved  with  initial conditions $\{\bar{x}(0)\}=
\{x^0\}, \{\bar{y}(0)\}= \{y^0\}$. Next, we are going  to  calculate  the
conditional   probability   distribution   $f(\{x\},\{y\},  t\mid\{x^0\},
\{y^0\},0)$ or $Q(\{\zeta\}, \{\xi\}, t\mid0,0,0)$.

Assuming  the variation of diffusion coefficient over the narrow width of
the distribution function at any instant of time to be $\bigcirc(D)$,  we
could replace the second Fokker-Planck coefficient $D(y)$ by $D(\bar{y})$
at  each instant of time. This assumption makes the calculation extremely
simple. Collecting coefficients $\bigcirc(D^0)$, we get back  quasilinear
Fokker-Planck equation for $Q$:

\begin{equation}
\label{eq.7}
\frac{\partial Q}{\partial t}+\sum_i(\frac{\dot{D}(\bar{y}_i)}
{D(\bar{y}_i)})Q
=-\sum_i[a_i\frac{\partial(\zeta_i Q)}{\partial\zeta_i}
        +b_i\frac{\partial(\xi_i Q)}{\partial\zeta_i}
        +c_i\frac{\partial(\xi_i Q)}{\partial\xi_i}
        +d_i\frac{\partial(\zeta_i Q)}{\partial\xi_i}
        -(1/2)\frac{\partial^2Q}{\partial\zeta_i^2}]
\end{equation}

where $a_i,b_i,c_i,d_i$ are given by
\begin{mathletters}
\begin{eqnarray}
\label{eq.8}
a_i&=&(\frac{\partial  h}{\partial\bar{x}_i})-  (\frac{\dot{D}(\bar{y}_i)
}{2D(\bar{y}_i)})\\
b_i&=&(\frac{\partial h}{\partial\bar{y}_i})\\
c_i&=&(\frac{\partial H}{\partial\bar{y}_i}) - (\frac{\dot{D}(\bar{y}_i)}
{2D(\bar{y}_i)})\\
d_i&=&(\frac{\partial H}{\partial\bar{x}_i})
\end{eqnarray}
\end{mathletters}

Eq.(2.7) suggests that
\begin{equation}
\label{eq.9}
Q(\{\zeta\},\{\xi\},t) = \prod_jQ_j(\zeta_j,\xi_j,t)
\end{equation}

where  the distribution function $Q_j$ for each $j$ satisfies the similar
equation written below without the subscript:

\begin{equation}
\label{eq.10}
\frac{\partial Q}{\partial t}+(\frac{\dot{D}(\bar{y})}
{D(\bar{y})})Q
=-[a\frac{\partial(\zeta Q)}{\partial\zeta}
        +b\frac{\partial(\xi Q)}{\partial\zeta}
        +c\frac{\partial(\xi Q)}{\partial\xi}
        +d\frac{\partial(\zeta Q)}{\partial\xi}
        -(1/2)\frac{\partial^2Q}{\partial\zeta^2}]
\end{equation}

subject to the initial condition

\begin{equation}
\label{eq.11}
Q(\zeta,\xi,t=0) = \delta(\zeta)\delta(\xi)
\end{equation}

The solution of Eq.(2.10) is given by

\begin{equation}
\label{eq.12}
Q(\zeta,\xi,t)  =  [\frac{1}{(2\pi)^2}]\int\int  exp-\{ik\zeta+il\xi  +
\frac{ [g(t)k^2+G(t)l^2 + 2C(t)kl]}{2D(t)}\} dk dl
\end{equation}

where   $g(t), \ G(t), \ C(t)$   satisfy   the   set  of  coupled  first  order
differential equations :

\begin{mathletters}
\begin{eqnarray}
\label{eq.13}
\frac{\dot{g}}{2}&  = &(\frac{\partial h}{\partial x})g + (\frac{\partial
h}{\partial    y})C+\frac{D}{2}\\
\frac{\dot{G}}{2}&=&(\frac{\partial
H}{\partial y})G+(\frac{\partial H}{\partial x})C\\
 \dot{C}  &=&(\frac{\partial  h}  {\partial  x})C  +  (\frac{\partial  h}
{\partial y})G+ (\frac{\partial H}{\partial  y})C  +  (\frac{\partial  H}
{\partial x})g
\end{eqnarray}
\end{mathletters}

with the initial conditions

\begin{equation}
\label{eq.14}
g(0)=G(0)=C(0)=0
\end{equation}

Once  $Q(\zeta,\xi,t)$  is  known,  from  Eq.(2.9)  and Eq.(2.5) the full
conditional   probability   distribution   function   $f(\{x\},    \{y\},
t\mid\{x^0\},  \{y^0\},0)$  is  known. Integrating this function over all
variables except one, say $x_i$, one identifies $g_i(t)$ as the  variance
of the stochastic variable $X_i$.

\begin{equation}
\label{eq.15}
<(X_i-<X_i>)^2> = g_i(t)
\end{equation}

We  note  that  the  homogeniety of Eq.(2.10) suggests that $<\zeta(t)> =
<\xi(t)>=0$, or the average of the variables $X$ and $Y$ at any time will
be  determined  by  their  deterministic  values  obtained   by   solving
Euler-Lagrange equation (2.6). Similarly, one observes from Eq.(2.12),

\begin{mathletters}
\begin{eqnarray}
\label{eq.16}
<(X_i-<X_i>)(Y_i-<Y_i>)>&=& C_i(t) \\
<(Y_i-<Y_i>)^2>&=& G_i(t)
\end{eqnarray}
\end{mathletters}

\subsection{Application to the fission process}

In  the fission process, in accordance with our previous work \cite{dha},
the shape of  the  fissioning  nucleus  is  described  in  terms  of  the
elongation axis (the neck parameter of \cite{bra} taken
equal to zero). Thus, in the dynamical description we have the elongation
axis, its relative orientation with respect to  an  inertial  system  and
respective  velocities  associated  with them as the stochastic variables
interacting with a large number of internal nucleonic degrees of  freedom
constituting  a heat bath at temperature $T$ determined by the excitation
energy available to it. We further assume  that  the  'collisional'  time
scale  of  the nucleonic degrees of freedom is much shorter than the time
scale of the macroscopic evolution of the collective variable so that  at
each  instant  of time the heat bath is assumed to be in quasi-stationary
equilibrium.

The  Euler-Lagrange  equations  (2.6)  were  solved  in our earlier works
\cite{dha}. To avoid repetition we deliberately omit  the  procedure  and
scheme  to  solve those equations. For the sake of completeness we merely
write those equations and refer to our previous  papers  to  clarify  the
details.

Giving correspondence to the terminology used in this paper, we associate

\begin{mathletters}
\begin{eqnarray}
Y_1=r, X_1=\dot{r},\\
Y_2=\theta, X_2=\dot{\theta}.
\end{eqnarray}
\end{mathletters}

Thus we have

\begin{mathletters}
\begin{eqnarray}
\label{eq.18}
H_1(\{x,y\}) &=& x_1 =\dot{r}\\
h_1(\{r,\dot{r}\})&    =    &[\frac{L^2}{\mu r^3} - \gamma\dot{r}    -
\frac{\partial(V_C+V_N)} {\partial r}]/\mu, \\
H_2(\{x,y\}) &=& x_2 =\dot{\theta},\\
h_2(\{\theta,\dot{\theta}\})& = &-(I_1 \ddot{\theta_1} + I_2 \ddot{\theta_2})
/I,\\
I_1 \ddot{\theta_1}&=&\gamma_t[g_2(\dot{\theta_2} - \dot{\theta}) +
g_1(\dot{\theta_1} - \dot{\theta})] g_1,\\
I_2 \ddot{\theta_2}&=&\gamma_t[g_2(\dot{\theta_2} - \dot{\theta}) +
g_1(\dot{\theta_1} - \dot{\theta})] g_2.
\end{eqnarray}
\end{mathletters}

The quantities $V_C$, $V_N$ represent the Coulomb and nuclear interaction
potentials  and  $\gamma$,  $\gamma_t$  are  the  radial  and  tangential
components of friction, respectively. The nuclear part of the interaction
is approximated by the proximity interaction \cite{vnuc}.
$I_1,  I_2$  are  the  moments  of
inertia of the two lobes and $L$ refers to the relative angular momentum.
$g_1$ and $g_2$ are the distances of the centres of mass of the two lobes
from  the  centre  of mass of the composite dinuclear system and the term
$[g_2(\dot{\theta_2}   -    \dot{\theta})    +    g_1(\dot{\theta_1}    -
\dot{\theta})]$  represents  the  relative tangential velocity of the two
lobes. The quantities $I$ and $\mu$ are the moment  of  inertia  and  the
reduced  mass  associated  with  the fissioning liquid drop, respectively
\cite{dha2}.

It  has  already  been  shown that the
tangential friction which causes dissipation of relative angular momentum
$L$ into the angular momenta $I_1,I_2$ of the two fragments does not have
any  significant  effect  on  the  physical  observables \cite{dha} .
Besides,   no
experimental  observation  of  angular  momentum  dispersion  of  fission
fragments are available in the litterature. Therefore, in  the  following
the calculations of higher moments are restricted to the radial degree of
freedom  only.  For the sake of convenience we omit the subscripts in the
functions $H$ and $h$ below.

The  variances  are obtained by solving Eqs.(2.13). The
diffusion coefficient $D$ is evaluated employing  Einstein's  fluctuation
dissipation theorem. Thus, the Eqs.(2.13) now becomes,

\begin{mathletters}
\begin{eqnarray}
\dot{g}(t)&=&2(\frac{\partial h_1}{\partial \dot{r}})g(t)+
2(\frac{\partial h_1}{\partial r})C(t)+
2\gamma(r)T(r)/ \mu^2\\
\dot{G}(t)&=& 2C(t)\\
\dot{C}(t)&=&(\frac{\partial h_1}{\partial \dot{r}})C(t)+
2(\frac{\partial h_1}{\partial r})G(t)+g(t)
\label{eq.19}
\end{eqnarray}
\end{mathletters}

with  the  initial  conditions(2.14).  The  initial conditions of $r$ and
$\dot{r}$ for solving Eq.(2.18) are\cite{dha}

\begin{equation}
\label{eq.20}
r_0=r(t=0)=r_{min} \pm \delta r_0, \ \dot{r}_0=\dot{r}(t=0)
=(\frac{E^*_0R_N}{2\mu})^{1/2}
\end{equation}

where  the  potential energy surface around the minimum is approximated
as harmonic oscillator with $\omega$ being the oscillator frequency.
At each instant of time, it is assumed that the state of the nucleus is
ameanable to a thermodynamic description with temperature $T$. Therefore,
$\delta r_0$ in Eqn. (2.20), which is taken as the root of the thermal
average of mean quantum dispersion around the minimum of the potential,
is expressed as $\delta r_0 = (\frac{\hbar}{2\omega\mu}\coth \frac
{\hbar\omega}{2T(r=r_{min})})^{1/2}$.
The quantity $R_N$ is a random number  between  0  and  1
from uniform probability distribution and $E^*_0$ is the initial
available energy. The temperature $T(r)$ in Eqn. (2.19a) has been
calculated from the instantaneous excitation energy $E^*(r)$ using the
relation $T(r) = \sqrt (E^*(r)/a)$ with $a = A/10$. At each instant, the
dissipated energy is added to, and the energy carried away by the
prescission particles (if any) is subtracted from, the excitation energy
which is then used to calculate the temperature at the next instant.

Solving  Eq.(2.18)  and Eqs.(2.19) simultaneously with initial conditions
(2.14) and (2.20) we generate the  conditional  probability  distribution
function  $f(r,\dot{r},t\mid  r(t=0),  \dot{r}(t=0), 0)$. The probability
distribution function $f(r,\dot{r},t)$ could be obtained as

\begin{equation}
\label{eq.21}
f(r,\dot{r},t) = \int f(r,\dot{r},t\mid r(t=0),\dot{r}(t=0),0)
                      f(r(t=0),\dot{r}(t=0),0)
                      dr(t=0)d\dot{r}(t=0)
\end{equation}

where  $f(r(t=0),\dot{r}(t=0),  0)$  is  the  probability distribution of
position and velocity of the stochastic variables at the initial time. As
described by the initial condition(2.20), this can be represented as

\begin{equation}
\label{eq.22}
f(r(t=0), \dot{r}(t=0), 0) = \delta(r(t=0)-r_{min}\mp \delta r_0)\times
f(\dot{r}(t=0))
\end{equation}

Here, we assumed that each fissioning nucleus in the ensemble starts from
a  fixed  initial  position  but  with  different  partioning  of initial
excitation  energy\cite{dha}.  Finally,  substitution  of  Eq.(2.22)   in
Eq.(2.21) would give

\begin{equation}
\label{eq.23}
f(r,\dot{r},t)        =        \sum_{R_N}f(r,\dot{r},t\mid       r_{min},
(\frac{E^*_0R_N}{2\mu})^{1/2},0)
\end{equation}

The  asymptotic  expansion  in  our  model  thus  provides  the following
picture: In the zeroth order approximation, the motion  is  described  by
the  Euler  -  Lagrange equation. This requires the initial momentum as a
generator of motion, which is supplied by a random  fraction  of  initial
available  energy $ E^*_0$ of the system. This initial randomness restricts
the trajectories to have fission fate thus providing the crosssection  of
the  residue.  The  first  order  approximation provides mostly the other
transport property, namely the variance of the physical variable. As  the
approximation  of  the  distribution function is over the solution of the
Euler - Lagrange equation of motion, this part provides  the  observables
of  the  escape  part  of the distribution. In this way, this model could
describe the bifurcation of the total distribution function if one  would
solve the full Fokker - Planck equation or the Langevin equation.

\section{Numerical calculation and results}
The applicability of the generalised formalism developed in Sec.~\ref{f-p}
has been tested  quite rigorously by confronting it with a wide range of
experimental data on various physical observables of the fission process.
The details of such calculation procedure has been reported elsewhere
(Ref.~\cite{dha}), and is given here in brief.
 
\subsection{The shape, friction and dynamics of the fissioning system}
\label{sec.fric}
For  the  present  calculation,  instantaneous  shape  of  the fissioning
nucleus is taken to be of the form \cite{dha,bra},

\begin{equation}
\label{eq.1r}
\rho^2(z) = c^{-2}(c^2-z^2)(A+Bz^2+\alpha z c),
\end{equation}

where  the  coefficients $A$ and $B$ are given by, $A=c^{-1}-Bc^2/5$, and
$B=(c-1)/2$, respectively. The variable $c$ corresponds to the elongation
and $\alpha$ is a parameter which depends upon the asymmetry  ($a_{asy}$)
defined as $a_{asy}={(A_1 - A_2)}/A_{CN}$, where $A_{CN}$ is the compound
nucleus  mass,  and  $A_1,  \  A_2$  correspond  to the masses of the two
fragments. The parameter $\alpha$ is related to the  asymmetry  $a_{asy}$
through  the  relation  $\alpha  =  .11937  a_{asy}^2  +  .24720 a_{asy}$
\cite{dha}. As the shape changes gradually, the coordinates  of  the  two
maxima  and  that of the minimum of the surface Eq.~(\ref{eq.1r}) change.
The scission point is defined when the minimum point touches the $z$-axis
and it is given by $A -c^2\alpha^2/4 B  =  0$  .  Therefore,  the
value  of  $c$  at  which  scission  occurs  depends  on $\alpha$ and the
dependence is given by $c_{sc} = -2.0 \alpha^2 + .032 \alpha + 2.0917$.

The  variable $r$ is defined as the centre to centre distance between the
two lobes. From the generalised shape  given  by  Eqn.~(\ref{eq.1r}),  we
first  construct  the  centres  of mass of left and right lobes, and call
them $z_l$ and $z_r$ respectively. Then $r$ is defined as $r  =|z_l-z_r|$
.  The  reduced  mass  parameter  $\mu$,  is obtained from the calculated
masses of the two lobes.

The  temporal  evolution of shape of the fissioning nucleus is assumed to
start from the minimum of the potential energy surface eventually leading
to scission.  The  fission  trajectories  are  obtained  by  solving  the
Euler-Lagrange  equations  with  conservative  forces  derived  from  the
nuclear   and   Coulomb   potentials   \cite{dha,dha2,dha3}.   For    the
non-conservative  part of the interaction, we would consider viscous drag
arising not only due to two body collision but also due to the collisions
of the nucleons with the wall or surface of the nucleus.  Hence  $\gamma$
in  Eqn.(2.18b)  contains two parts; $\gamma^{TB}$ and $\gamma^{OB}$, for
two-body and one-body dissipative mechanisms, respectively. Assuming  the
nucleus  as  an incompressible viscous fluid, and for nearly irrotational
hydrodynamical  flow,  $\gamma^{TB}$  is  calculated  by   use   of   the
Werner-Wheeler method \cite{nix,feld} and is given by

\begin{mathletters}
\label{eq.10r}
\begin{equation}
\gamma^{TB} = \pi \mu_0 \ R_{CN} h(\alpha)
f(\frac{\partial c}{\partial x})
\int^{+c}_{-c} dz\rho^2[3A^{' 2}_c +\frac{1}{8}\rho^2A^{''2}_c]
\end{equation}

where the factor $h(\alpha) = \exp(- K \alpha^2)$ is included in order
to explain the observed fragment asymmetry dependence of neutron
multiplicity (for details, see Ref.~\cite{dha}), and,

\begin{equation}
\label{eq.11r}
A_c(z) = -\frac{1}{\rho^2(z)} \frac{\partial}{\partial c}
\int^z_{-c} dz'\rho^2(z').
\end{equation}
\end{mathletters}

The  quantities  $A'_c,A''_c$  are  the  first  and second derivatives of
$A_c(z)$  with  respect  to  $z$.  $\mu_0$  is  the  two  body  viscosity
coefficient. The factor $f(\frac{\partial c}{\partial x})$ is taken to be

\begin{equation}
\label{eq.12r}
f(\frac{\partial c}{\partial x}) = (\frac{\partial c}{\partial x})^2 +
2(\frac{\partial c}{\partial x }),
\end{equation}

where $x=r/R_{CN}$, $R_{CN}$ being the radius of the compound nucleus.

 The  tangential friction $\gamma_t^{TB}$ is calculated using the following
relation \cite{dha},

\begin{equation}
\label{g_tb}
\gamma_t^{TB} = (\frac{\partial c}{\partial n})^2 \gamma_r^{TB},
\end{equation}

where $n$ (the value of $\rho$ at the minima of $\rho^2$ in Eqn.~\ref{eq.1r})
is the instantaneous neck radius of the fissioning system.

One  body  dissipative  force,  $F_{dis}$,  is  obtained from the rate of
energy dissipation, $E_{dis}$, by

\begin{equation}
\label{eq.15r}
F_{dis} = -\frac{\partial}{\partial \dot{x}} E_{dis}(x)
\end{equation}

where  $\dot{x}$ refers to the rate of change of $x$ with respect to time
and $E_{dis}(x)$ is the rate of energy dissipation at $x$ given by

\begin{equation}
\label{eq.16r}
E_{dis} = \frac{1}{2} \rho_m \bar{v} \oint dS \ {\dot{\vec{e}_n}}^2,
\end{equation}

where  $\vec{e}_n$  is  the  unit  normal  direction  at the surface. The
integration is done over the  whole  surface.  $\rho_m$  is  the  nuclear
density  and  $\bar{v}$  is the average nucleonic speed obtained from the
formula

\begin{equation}
\label{eq.17r}
\bar{v} = \surd(\frac{8k}{m\pi}) (E_{av}/a)^{1/4}
\end{equation}

with  $E_{av}$  is  the available energy and the level density parameter,
$a$, is taken to be $A_{CN}/10$. For the generalised shape (\ref{eq.1r}),
one-body friction, $\gamma^{OB}$, is obtained as

\begin{equation}
\label{eq.18r}
\gamma^{OB} = 2\pi \ \rho_m \bar{v} R_{CN}^2  f(\partial c/\partial x)
\int^{+c}_{-c} dz \ \rho \ [1+ \ \rho\prime^2]^{-1/2}
[A_c\rho\prime + (1/2) \ \rho A'_c]^2
\end{equation}

where  $\rho\prime,  \  A'_c$  are  the  derivatives  of $\rho ,A_c$ with
respect to $z$ and all other quantities are defined earlier. The tangential
part of the one-body friction is  calculated in a similar manner as
in Eqn.~\ref{g_tb}.

The  friction  forces  used  in  the  calculation  are  taken  as follows
\cite{dha}. One-body 'wall' friction has been used in the ground state to
saddle region, where nuclear shapes are nearly mononuclear. The  strength
of  the  one-body  friction  used  was attenuated to 10\% of the original
'wall' value. This weakening of the wall friction has also been confirmed
from the study of the role  of  chaos  in  dissipative  nuclear  dynamics
\cite{spal}.  In  the  saddle  to scission region, on the other hand, the
nuclear  dissipation  was taken to be of two-body origin and the value of
the  viscosity coefficient $\mu_0$ used in the present calculation was (4
$\times10^ {-23} MeV \cdot sec \cdot fm^{-3}$).  This  value  of  $\mu_0$
corresponds  to  0.06  TP   ($1  TP=6.24\times10^{-22}MeV\cdot sec \cdot
fm^{-3}$).

\subsection{Prescission neutron emission}
\label{sec_npre}

\subsubsection{Prescission Neutron Multiplicities}
\label{sec_nmul}

The  emission  of  the  prescission neutrons is simulated in the 
following way. During the temporal evolution of the  fission  trajectory
the  intrinsic  excitation  of  the system, and {\it vis-a-vis},
the neutron decay width  at  each  instant,$\Gamma_n$,  is
calculated  using the relation $\Gamma_n = \hbar W_n$. The decay rate
$W_n$ is given by,

\begin{equation}
\label{eq.wn}
W_n = \int^{E_{max}}_0 dE \frac{d^2\Pi_n}{dEdt},
\end{equation}

where, $d^2\Pi_n/dEdt$ is the rate of decay $A\rightarrow A-1+n$
in an energy interval $[E,E+dE]$ and a time interval $[t,t+dt]$.
The quantity $d^2\Pi_n/dEdt$ may be evaluated using standard
expression \cite{dha}. 

The emission of neutrons during the temporal
evolution of the trajectory is simulated as follows. At each time step, 
the probability of emission of a neutron, $\tau/\tau_n$ (where 
$\tau_n(=\hbar/\Gamma_n)$, $\tau$ are the neutron decay time and the time
step of the calculation, respectively),  is computed and compared with a
random  number $R_N$ from a uniform probability distribution. The
emission of a neutron is assumed to take place, if it satisfies the
following criterion \cite{dha};

\begin{equation}
\label{eq.25}
\tau/\tau_n > R_N.
\end{equation}

If the condition (\ref{eq.25}) is not
satisfied, no emission of neutron takes place. The time step $\tau$ is
chosen in such a way that it satisfies the condition $\tau/\tau_n  \ll  1$.
Consequently  the probability  of  emission  of  two or more neutrons in
time $\tau$ would be extremely small.
The calculation is continued over the whole trajectory for a number of
times at each angular momentum $\ell$ to estimate the average prescission
multiplicity at each $\ell$. The calculation is then repeated for all allowed
values of angular momentum to compute the aveage value of prescission
neutron multiplicity  $n_{pre}$ \cite{dha}.

\begin{figure}
\centering
\epsfysize=10cm
\epsffile[13 13 599 780]{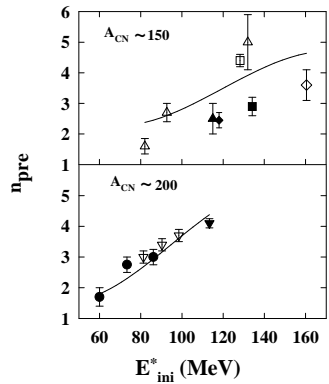}

\vspace{-1cm}

\caption{Prescission  neutron  multiplicities  plotted as a function of the
initial excitation energy $E^*_{ini}$ of the compound nuclei of  masses
A$_{CN}  \sim$
150 ({\it upper half}), and A$_{CN} \sim$ 200 ({\it lower half}). The solid
curve is the present calculation. Different symbols correspond to different
sets   of   experimental  data,  ({\it  ie,}  filled  circle
\protect\cite{e4},    open    inverted     triangle
\protect\cite{e4},   filled  inverted
triangle \protect\cite{e3}, open triangle
\protect\cite{e4},   filled   triangle
\protect\cite{e6},   open   diamond
\protect\cite{e6},   filled   diamond
\protect\cite{e3},    open   square
\protect\cite{e3},   filled    square
\protect\cite{e3}).}

\label{fig_mul}
\end{figure}

The  calculated  values of $n_{pre}$ have been displayed in
Fig.~\ref{fig_mul} alongwith the respective experimental data
as a function of the initial excitation  energy  of  the
compound nucleus for two different mass regions. The
solid  curves  are  the  results  of  the  present  calculations  and
the  symbols  correspond  to  experimental  data
\cite{e3,e4,e6}. It is seen that for heavier systems ($A_{CN}  \sim  200$)
({\it lower half}),
the  theoretical  predictions  are in good agreement with the corresponding
experimental  data.  For  lighter  systems   ($A_{CN} \sim 150$)
({\it upper half}), the  experimental values of $n_{pre}$ have larger
uncertainties and fluctuations, and the  theory is seen to
reproduce quite well the average trend of the data.
A part of this fluctuation  in neutron  emission  here
may be due  to specific structure effects of different
compound systems; for example, $^{162}Yb$ ({\it  filled diamond})
is quite neutron deficient compared to $^{168}Yb$ ({\it  open triangle}),
and  neutron emission from the former is therefore expected to be
somewhat less. Similarly, at high incident energies ($>$ 10 MeV/nucleon)
the  observed  multiplicity ({\it open diamond}) was found to be lower than
the average theoretical trend, which may be due to 
the noninclusion of the effect of preequilibrium emission in the present
calculation.

\begin{figure}
\centering
\epsfysize=10cm
\epsffile[13 13 599 780]{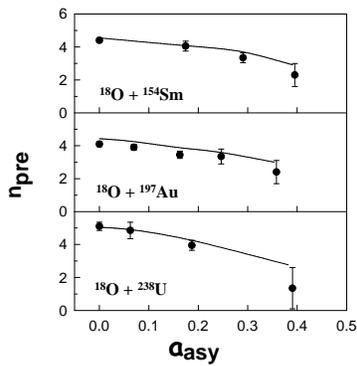}

\vspace{-1cm}

\caption{Prescission  neutron  multiplicity  $n_{pre}$  as  a  function  of
fragment  mass  asymmetry  $a_{asy}  $for  $^{18}O$  induced  reactions  on
$^{154}Sm,  \  ^{197}Au  \ \rm{and} \ ^{238}U$. Filled circles correpond to
the experimental data \protect\cite{e3} and
the solid curves are the present calculations.}

\label{fig_asym}
\end{figure}

The  fragment mass asymmetry dependence of neutron multiplicity is displayed
in  Fig.\ref{fig_asym} for 
$^{18}O \  (E_{lab}  =  158.8  \  MeV)$  induced  reactions  on
$^{154}Sm$,  $^{197}Au$  and  $^{238}U$ \cite{e3}.
The solid circles correspond to the  experimental data and the solid lines
are the theoretical predictions of the same. It is found that the present
calculations agree quite well with the experimental data in all the cases.
The value of the constant $K$ (Eqn.~\ref{eq.10r}) was found to be
$161\pm3$  which  is independent of the mass of the compound system. It is,
therefore, interesting to note that with the inclusion of the  term
$h(\alpha)$ in the friction form factor (Eqn.~\ref{eq.10r}),  we are able
to  explain the  prescission  neutron
multiplicity data for both symmetric as well as asymmetric fission
with the same value of the viscosity coefficient, $\mu_0$
(= $4 \times10^ {-23} MeV \cdot sec \cdot fm^{-3}$).

\subsubsection{Energy of emitted neutrons}
\label{sec_tke}

The  kinetic  energy  of  the  emitted  neutron is extracted through random
sampling technique \cite{dha}. Assuming that the system  is  in
thermal  equilibrium at each instant of time $t$, the energy
distribution of the emitted neutrons is represented  by  a  normalised
Boltzmann  distribution  corresponding  to the instantaneous temperature of
the system. From a uniformly distributed random number  sequence  \{$x_n$\}
in  the  interval  [0,1],   another  random  number  sequence \{$y_n$\}
with  probability  distribution   $f(y)$ is constructed, where $f(y) \sim
\exp(-\beta(t)  y)$ is a normalised Boltzmann distribution corresponding to
the temperature $\beta(t)$ at any instant of time $t$. Then,  the  sequence
\{$y_n$\} is obtained from the sequence \{$x_n$\} by the relation,

\begin{equation}
y(x) = F^{-1}(x).
\end{equation}

Here, $F^{-1}$ is the inverse of the function $F(y) = x = \int_0^y f(y)
dy $, which is computed numerically by forming a table of integral values.
The energy of the emitted neutron is given by $E_n = y E_n^{max}$,
where $E_n^{max}$ is chosen in such a way that the Boltzmann probability at
that energy is negligible for all instants of time $t$. After the  emission
of  the  neutron,  the  intrinsic excitation energy is recalculated and the
trajectory is continued.

\begin{figure}
\centering
\epsfysize=10cm
\epsffile[13 13 599 780]{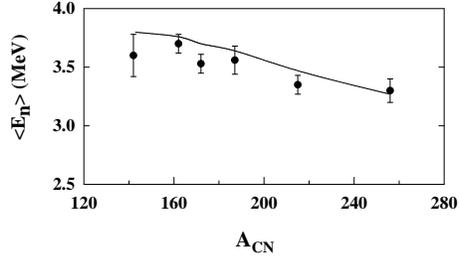}

\vspace{-1cm}

\caption{mean
energy of the evaporated neutrons ($<E_n>$)
plotted as a function of A$_{CN}$. The solid curves are
the present calculations, and the filled circles are the corresponding data
\protect\cite{e3}.}

\label{fig_ke}
\end{figure}

The  average  energy  of  the
prescission  neutrons,  $<E_n>$  
has been plotted as a function of  the  compound  nuclear  mass,
A$_{CN}$, in Fig.~\ref{fig_ke}. It is
seen  from  the  figure  that  the  theoretical  predictions of $<E_n>$ 
({\it solid curve}) are in good agreement with  the  respective
experimental  data  ({\it  filled  circles}).

\subsection{Average and variance of TKE}

\begin{figure}
\centering
\epsfysize=10cm
\epsffile[13 13 599 780]{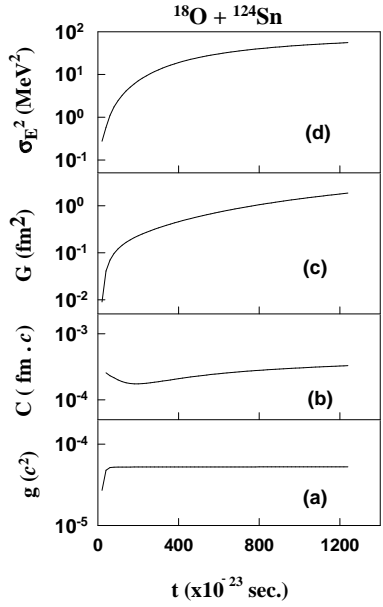}

\vspace{-1cm}

\caption{Variation  of  {\it  (a)}  $g(t)$,  {\it  (b)} $C(t)$, {\it (c)}
$G(t)$  and  {\it  (d)}  $\sigma^2_E$  as  a function of time $t$ for the
system $^{16}$O+$^{124}$Sn.}
\label{fig1}
\end{figure}

The temporal evolutions of the variables $g(t)$, $C(t)$, $G(t)$ along the
fission  trajectory  have  been  computed  for  a  representative  system
$^{16}$O  + $^{124}$Sn and the results are plotted in Fig.~\ref{fig1}. It
is seen from the figure that, initially, all of them increase steeply and
then their magnitudes become nearly constant throughout the rest  of  the
trajectory.  Furthermore,  the calculation shows that $C^2(t)/g(t)G(t)\ll
1$, which implies that the correlation of position and  velocity  of  the
elongation  variable  $(r)$  is much smaller compared to their respective
variances. The variance of  energy  and  average  of  total kinetic
energy (TKE)  at scission point are given by,

\begin{mathletters}
\begin{eqnarray}
\sigma_E^2  &=&  (\mu\dot{r})^2g(t) + [{\partial(V_C+V_N)} / {\partial
r}]^2G(t), \label{sige2a}\\
<E(t)> &=& \mu g(t_{sc})/2+E_{det}. \label{sige2b}
\end{eqnarray}
\end{mathletters}

The     contribution     of    term    (typically    $\sim    2\mu\dot{r}
({\partial(V_C+V_N)} / {\partial r})C(t)$  )  involving  the  correlation
between  position  and  velocity  has been neglected in Eq.~(\ref{sige2a})
as it is
quite small compared  to  the  other  terms  invoving  the  variances  of
position  and  velocity.  The  quantity  $t_{sc}$ is the time at scission
point  and $E_{det}$ is the deterministic value of total fragment kinetic
energy (TKE) after scission and $\sim$~ 100 - 200 MeV. It is assumed that
the variation of the potential over the narrow width of  the  probability
distribution   is   small  so  that  the  average  of  the  potential  is
approximated as the value of the potential  at  the  mean  position.  The
variation  of  the  kinetic energy variance $\sigma_E^2$ as a function of
time  has  also  been  displayed  in  Fig.~\ref{fig1}.   The   value   of
$\sigma_E^2$  is  also seen to increase steeply at the beginning and then
it becomes nearly constant throughout the rest of the time. As  envisaged
earlier,  the  result  clearly  shows  that ${\sigma_E}/{<E>}\ll1$, which
demonstrates the validity of asymptotic expansion in deriving the  result
instead of solving the Fokker-Planck equation in detail.

\begin{figure}
\centering
\epsfysize=10cm
\epsffile[13 13 599 780]{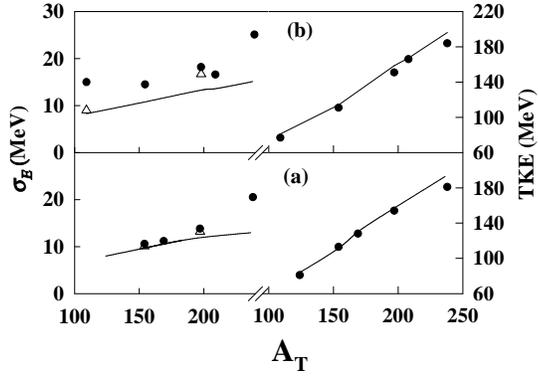}

\vspace{-1cm}

\caption{ Variation of $\sigma_E$ and mean TKE as function of target mass
number  $A_T$,  for  {\it (a)} 158.8 MeV $^{18}$O , and {\it (b)} 288 MeV
$^{16}$O induced fission reactions.  Filled  circles  correspond  to  the
experimental data \protect \cite{e3} and solid curves are the
present  theoretical results. Open triangles are the values of $\sigma_E$
obtained using Eqs. (3.13) {\it (see text).}}
\label{fig2}
\end{figure}

The  theoretical  predictions  of  $\sigma_E(th)$  and mean total kinetic
energy (TKE) (solid curves) for the fission of several compound systems
produced in  the 158.8 MeV $^{18}$O, 288 MeV $^{16}$O  induced
reactions on various targets have
been displayed in Figs.~\ref{fig2}a,~\ref{fig2}b, respectively, alongwith
the experimental  data  \cite{e3}  (filled circles). From the figure it is
observed that theoretical predictions of TKE agree quite  well  with  the
respective experimental data for all the systems studied. However, it may
be  noted  here,  that  the  experimental  values  of $\sigma_E(exp)$ are
usually  obtained  by  averaging  over  the  full  mass  yield  spectrum.
Therefore,  $\sigma_E(exp)$  consists of two terms, {\it viz.}, ({\it i})
contributions arising due to stochastic fluctuations in the  dynamics  of
fission   process,$\sigma_E$,  and  ({\it  ii})  contributions  from  the
variation of the mean kinetic energy with the  fragment  mass  asymmetry,
$\sigma_E(kin)$. So, $\sigma_E(exp)$ may be written as \cite{laza},

\begin{mathletters}
\begin{eqnarray}
\sigma_E^2(exp)& = &\sigma_E^2 + \sigma_E^2(kin), \\
\sigma_E^2 &=&\sum_{A_1}  \sigma_E^2(A_1, A_2) \cdot Y(A_1),\\
\sigma_E^2(kin) &=&\sum_{A_1}[ \bar{E} - <E(A_1,A_2)> ]^2  \cdot Y(A_1).
\end{eqnarray}
\end{mathletters}

Here $\sigma_E^2(A_1, A_2)$ and $<E(A_1,A_2)>$ are the variances and mean
values  of  the  total  kinetic energy of two fission fragments with mass
numbers $A_1$ and $A_2$ (compound nucleus mass $ A_{CN} =  A_1  +  A_2)$,
$\bar{E}$  being  the  average  of  $<E(A_1,A_2)>$  over  the  normalised
fragment mass yield, $Y(A_1)$ with $\sum_{A_1} Y(A_1)  =  1$.  Thus,  the
calculated value of $\sigma_E^2(th)$ is then compared with the stochastic
component  of  the experimental variance, {\it i.e.,} $\sigma_E^2$, which
is obtained after substracting $\sigma_E^2(kin)$  from  $\sigma_E^2(exp)$
(as mentioned above).

 We  have  extracted  $\sigma_E^2(kin)$  for  a few systems for which the
experimental fragment mass yield data are available \cite{e3},  taking
$<E(A_1,A_2)>$   from   Viola   systematics   \cite{viola}.   The  values
$\sigma_E$, i.e.,  $\sqrt{(\sigma_E^2(exp)  -  \sigma_E^2(kin))}$  ,  are
shown  in  Fig.\ref{fig2} as open triangles and they agree very well with
the  predicted  values  of  TKE variance. It is seen from Fig.~\ref{fig2}a
that  when  the projectile energy (and vis-a-vis the excitation energy of
the fused composite) is relatively lower,  the  calculated
values  are  in  fair  agreement  with the data. However, the calculation
underpredicts the experimental  value  of  $\sigma_E$  for  the  heaviest
target  considered  ($^{238}$U in the present case). With the increase in
the projectile energy (and the excitation energy of the composite), 
the  theoretical  predictions  are  found  to
underestimate  the  corresponding experimental values and the discrepancy
between the two increases with the increase in mass number
(Fig.~\ref{fig2}b),.

\begin{figure}
\centering
\epsfysize=10cm
\epsffile[13 13 599 780]{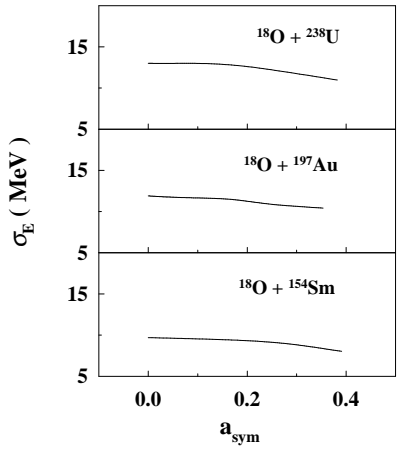}

\vspace{-1cm}

\caption{  Variation  of  predicted values of $\sigma_E$ as a function of
fragment  mass  asymmetry,  $a_{sym}  =   |A_1   -   A_2|/   A_{CN}   $.}
\label{sig_asym}
\end{figure}

 We  have  also  studied the fragment mass asymmetry dependence of energy
variance, $\sigma_E^2(A_1, A_2)$ for some representative systems and  the
results   are   displayed   in   Fig.~\ref{sig_asym}.  It  is  seen  from
Fig.~\ref{sig_asym} that the theoretical values of variances have only  a
weak dependence on the fragment mass asymmetry.

\section{summary and conclusions}

We  have  developed  a generalised formulation of asymptotic expansion of
the  Fokker-Planck  equation  for  the  systems   where   the   diffusion
coefficient  depends  on  the  stochastic  variable  explicitly. With the
assumption that the relative fluctuation of collective variable is  small
we  have  derived  the  equation  for  various  moments. The formalism is
applied to the case of fission where the  fluctuation  in  total  kinetic
energy  is  small  as  compared to its mean value. We have taken only one
degree of freedom, namely the elongation axis in  our  calculation.
%This
%constrains  the  fission trajectories over the potential surface with the
%value of the second parameter, $h$[8] (which describes the  variation  of
%the  thickness of the neck without changing the elongation length $ 2e $)
%, as zero.
However, one could incorporate neck degree  of  freedom  also in  a
more  realistic  calculation  based on the present formalism. The primary
motivation of the present work is  to  show  that  this  formalism  could
explain  the  basic features of the fission dynamics quite satisfactorily
without invoking the solution of  Fokker-  Planck  equation  or  Langevin
equation in detail.

The present model is found to explain  fairly well the observed neutron
multiplicities and their fragment mass asymmetry dependence as well as
the average energy of the evaporated neutrons over a wide range of mass
and excitation energies of the compound system with a single value of
the viscosity coefficient, $\mu_0$.
The  predicted  values of TKE are found to be in good agreement with the
experimental data and the theoretical estimates of the associated  TKE
variances  are  also found to  agree quite well with the respective numbers
extracted from the experimental data for the systems where the
fragment mass yield data  are  available.  For  a  more  direct  test  of
theoretical  models  it  is  necessary  that  experimental  estimation of
variances should not have admixture of other contributions arising due to
the variation of mean kinetic energy over different mass yields. This may
be achieved if measurements are done in smaller mass bins.

In  the  present studies, the correlation of the position and velocity of
the elongation axis has been found to be small.  However,  in  the  cases
where  such  condition  is  not  valid  the  energy variance still can be
calculated by adding  a  term  $2\mu\dot{r}({\partial(V_C+V_N)}/{\partial
r})C(t)$.  The  procedure  developed  here  could systematically generate
higher order hierarchies for relatively larger fluctuations than the ones
encountered in the present studies. In those cases one may have to  solve
the higher order equation which would involve higher order derivatives of
the functions $h(x,y)$ and$H(x,y)$, in general.

\end{document}